# The role of electron – hole pair formation in organic magnetoresistance

Sayani Majumdar,[1,2] [*] Himadri S. Majumdar,[1] Harri Aarnio,[1] Dirk Vanderzande[3], Reino Laiho[2] and Ronald Österbacka[1]

[1]Center for Functional Materials and Department of Physics, Åbo Akademi University, Turku, Finland.

[2]Wihuri Physical Laboratory, University of Turku, Turku, Finland.

[3] Chemistry Division, Institute of Material Research, Universiteit Hasselt, B3590 Diepenbeek, Belgium.

[*] *Corresponding author, e-mail: sayani.majumdar@utu.fi*

**Abstract**

Magneto-electrical measurements were performed on diodes and bulk heterojunction solar cells (BHSCs) to clarify the role of formation of coulombically bound electron-hole (e-h) pairs on the magnetoresistance (MR) response in organic thin film devices. BHSCs are suitable model systems because they effectively quench excitons but the probability of forming e-h pairs in them can be tuned over orders of magnitude by the choice of material and solvent in the blend. We have systematically varied the e-h recombination coefficients, which are directly proportional to the probability for the charge carriers to meet in space, and found that a reduced probability of electrons and holes meeting in space lead to disappearance of the MR. Our results clearly show that MR is a direct consequence of e-h pair formation. We also found that the MR line shape follows a power law-dependence of $B^{0.5}$ at higher fields.

Organic Magnetoresistance (OMAR) is an universal low-field magnetoresistance (MR) effect seen in organic diodes without magnetic electrodes [1,2]. Since the MR in organic materials is highest known among all non-magnetic materials, this effect can readily be used for magnetic sensors and magnetically controlled optoelectronic devices. This phenomenon, combined with cheap device manufacturing, raises a natural interest within the applied physics community. The mechanism governing OMAR is actively debated and three different models have been proposed to explain the OMAR effect. They are the magnetic field induced singlet triplet inter-conversion (MIST) model [3], the triplet-exciton polaron quenching model [4] and the bipolaron model [5]. The first two models have explained the OMAR effect to be due to the spin dependent electron hole pair (e-h) and, subsequent, exciton formation followed by exciton-exciton or exciton-charge interactions. The bipolaron model relies on the spin dependent formation of double occupancy of a particular site i.e. formation of bipolarons and subsequent enhancement or inhibition of carrier transport as the principal cause of OMAR. The main difference between the first two models and the bipolaron model is that the bipolaron model is a single-carrier model while the first two are dependent on singlet or triplet charge pair and exciton formation, e.g. a two-carrier process. Furthermore, Hu et al. [6], have shown that the MR in organic light emitting diodes (OLED) can be tuned between positive and negative values for the same device as explained using a double carrier process. They proposed that excited states in OLEDs either dissociate (singlets) or react (triplets) with charges to produce secondary charge carriers which modify the injection current accordingly to give either positive or negative MR. The magnetic field effect is caused by the inter-conversion between singlet and triplet states of the bound e-h pairs.

In low mobility materials ($\mu<1\text{cm}^2/\text{Vs}$) like disordered organic materials when the hopping distance is much shorter than the Coulomb radius $r_C = e^2/4\pi\varepsilon\varepsilon_0 kT$, where $e$ is the electronic charge, $\varepsilon(\varepsilon_0)$ is the relative (absolute) permittivity, $k$ is the Boltzmann constant and $T$ the temperature, the recombination of free electrons and holes into coulombically bound pairs follow the Langevin process [7]. The recombination coefficient is therefore proportional to the mobility and is described by the Langevin recombination coefficient $\beta_L = e(\mu_p + \mu_n)/\varepsilon\varepsilon_0$, where $\mu_p$ ($\mu_n$) is the hole (electron) mobility. The Langevin recombination can, therefore, be seen as the upper limit for the recombination efficiency in disordered systems. In the Langevin process the recombination is determined by the probability for the charge carriers to meet in space, independent of the subsequent fate of the formed e-h pairs. The formed e-h pairs can either dissociate back to free charge carriers or form excitons which later recombine either radiatively or non-radiatively.

Bulk heterojunction solar cells (BHSCs) are blends with a conjugated polymer donor and an acceptor molecule, typically a fullerene derivative 1-(3-methoxycarbonyl)propyl-1-phenyl-[6,6]-methanofullerene (PCBM). In BHSCs whenever an exciton is created it dissociates very fast, typically within ~ 40 fs [8] to form a coulombically bound e-h pair at the donor acceptor interface [9]. These e-h pairs can dissociate to free charges but the back-transfer to an excitonic state is energetically unfavorable. The lifetime of such a coulombically bound e-h pair depends on the relative position and spin-configuration and, in principle, the pair can stay bound for an extended period of time. However, in BHSCs made from regio-regular poly(3-hexylthiophene) (RRP3HT) mixed with PCBM in dichlorobenzene the dissociation of the coulombically bound e-h pairs is very effective and the probability for the charge carriers to meet in space (after photogeneration or injection) is drastically reduced. This is experimentally observed as a greatly reduced bimolecular recombination coefficient by as much as four orders of magnitude compared to the Langevin-recombination coefficient, i.e. $\beta/\beta_L \sim 10^{-4}$ [10]. The reason for this large reduction is not fully understood but the nano-scale morphology and, especially, the lamellar structure exhibited by RRP3HT seems to be very important parameters [11, 12]. However, in BHSCs made from blends of poly[2-methoxy-5-(3',7'-dimethyloctyloxyl)]-1,4-phenylene vinylene (MDMO-PPV) and PCBM, where the MDMO-PPV does not form 2D lamellar structures, the e-h pair dissociation is only around 60% [9] and the recombination is almost of Langevin type with $\beta/\beta_L \sim 0.5$ [10], even though the excitons are fully quenched. Since the probability for e-h pair formation is proportional to the normalized recombination coefficient ($\beta/\beta_L$) we therefore find BHSCs, with different recombination coefficients, as an ideal test system to clarify whether OMAR is caused by the formation of e-h pairs or it is a single carrier process. According to the bipolaron picture [5], a reduced e-h pair formation probability would cause an increase in the bipolaron formation probability, while according to the excitonic models [3,4] we should observe a weak OMAR in all solar cells due to the quenching of the excitons.

The device structure used in the experiment for the diode is indium tin oxide (ITO)/poly(3,4-ethylenedioxythiophene)-poly(styrenesulphonate)(PEDOT:PSS)/ RRP3HT/ lithium fluoride (LiF)/Al and for the solar cells (ITO/ PEDOT:PSS/ RRP3HT: PCBM or MDMO-PPV:PCBM/LiF/Al. The ITO coated glass electrodes were coated with a very thin layer of PEDOT:PSS and annealed at 120°C for 15 minutes. The π-conjugated polymer RRP3HT was spin-coated from a chloroform or dichlorobenzene solution and annealed at 120°C for 15 minutes. Finally the LiF and the aluminium electrode were vacuum evaporated to complete the device structure. For solar cells, a 1:1 blend of RRP3HT or MDMO-PPV and PCBM was used. The device preparation was done in a nitrogen-filled glove-box and using anhydrous solutions. After fabrication, the devices were transferred in nitrogen atmosphere to a cryostat placed in between the pole pieces of an electromagnet capable of producing up to 300 mT magnetic field. The resistance of the device is then measured in dark by sending a constant current through the device and measuring the voltage drop across it in a varying magnetic field in the temperature range 100 – 300 K. If not mentioned otherwise, the MR was measured at a current > 100μA. For measuring the recombination co-efficients, a 6 ns Nd:YAG laser operating in the second harmonic (532 nm) with the energy of 0.3 mJ per pulse is used. Integral mode of time of flight (TOF), charge extraction by linearly increasing voltage (CELIV) and double injection transients were used to measure the extracted charge carriers experimentally and from these measurements the $\beta/\beta_L$ ratio was calculated [10].

The MR, defined as

$$\%MR = \frac{R_B - R_0}{R_0} \times 100 = \frac{\Delta R_B}{R_0} \times 100 \qquad (1)$$

in the RRP3HT based diodes for a certain bias current at room temperature shows similar behavior as reported earlier [2]. Fig. 1(a) shows %MR as a function of magnetic fields for different measuring currents at room temperature. Positive MR up to 16% was found at room temperature when measured with 100 μA. However, we did not observe the reported universal MR line-shapes of either Lorentzian $\%MR(B) \sim B^2/(B_0^2+B^2)$ or a specific non-Lorentzian $\%MR(B) \sim B^2/(|B|+B_0)^2$ type [5]. The lineshape of the MR traces show distinct dependence of measuring current and a crossover from a Lorentzian line

shape at smaller magnetic fields and smaller current to a power law fit with $B^{0.5}$ dependence. Figure 1(b) shows the fitting of different line shapes for a typical RRP3HT diode. For lower measuring current (1 µA), the line shape is a Lorentzian until 80 mT, while for higher measuring currents (100 µA) the deviation from a Lorentzian line shape occurs at an even lower field. The magnetic field value $B´$, where the deviation occurs, decreases exponentially with increasing current. Similar line shape of magnetoconductance have also been observed earlier by Desai et al. [4] for Alq$_3$ based OLEDs where they found the OLED efficiency rising till $B´$ although the OMAR effect increases till 300 mT. These observations suggest that there are two different phenomena which govern the spin dynamics of the charge carriers in the diode devices and the entire line shape can not be fitted to a single function. The MR line shape in organic semiconductors is not necessarily universal and varies depending on operating conditions of the device. The dependence of MR on the square root of $B$ at higher magnetic fields (%MR ~$B^{0.5}$) is a well known phenomenon in colossal magnetoresistance materials and is indicative of a highly disordered electronic system with strong electron–electron interaction effects [13]. However, detailed temperature, higher magnetic field, material and thickness dependent study of the MR line-shapes are needed to clarify the effect.

Fig 2 shows the current–voltage and %MR–voltage characteristics for both a typical diode and BHSC measured between 0 and 5V bias in the dark and at room temperature. The MR increases rapidly beyond the threshold voltage ($V_{th}$) of the RRP3HT diode (marked by the arrow in the figure) but for the BHSCs negligible MR is observed - almost 3 orders lower in magnitude compared to the RRP3HT devices. In these BHSCs we have measured $\beta/\beta_L \sim 10^{-3}$. Thus, by reducing the probability of e-h pair formation in good solar cells, we have observed a dramatic decrease in the MR value. This observation is in good agreement with Lee et al., who found similar behavior in light-induced MR response on P3HT/PCBM solar cells [14].

To further investigate the dependence of e-h recombination on the OMAR response, we prepared BHSC devices with varying $\beta/\beta_L$ ratios and measured their magneto-transport properties. Figure 3 shows MR as a function of magnetic field for different devices exhibiting different $\beta/\beta_L$ ratios. The solar cells from RRP3HT:PCBM blend in dichlorobenzene have $\beta/\beta_L \sim 10^{-3}$ and showed the smallest OMAR - around $10^{-2}$ %. For BHSCs made from RRP3HT:PCBM blends in chloroform the $\beta/\beta_L$ ratio is $\sim 10^{-1}$ with a OMAR response of 0.2 % and for MDMO-PPV:PCBM devices $\beta/\beta_L \sim 0.5$ [10] showing almost as high MR response as the RRP3HT device alone. This means that even though excitons are completely quenched in all the BHSCs, we see an increase in the MR response when the probability to form e-h pairs is increased, as seen in the MDMO-PPV/PCBM devices.

The mechanism for the MR from coulombically bound e-h pairs is still debatable. While Frankevich et al [15] and Hu et al [6] conclude that spin-mixing of singlet and triplet e-h pairs under magnetic field is possible, recent results from McCamey et al. [16] show that spin mixing in the bound polaron pair state is highly improbable. Together with our results, it means that the MR is a direct consequence of the e-h pair *formation process*. However, further study is needed to clarify this issue.

In order to determine the role of balanced carrier injection on OMAR we prepared BHSCs without PEDOT:PSS on ITO substrates causing inefficient hole injecting interface and also devices without LiF on Al side having poor electron injecting contact.

In all these cases the normalized recombination coefficient was $\beta/\beta_L \sim 10^{-3}$. In Fig. 4(a) we show the *I-V* characteristics of the BHSCs with both good injecting and modified electrodes and the inset of Fig. 4(b) shows the band diagram of our BHSCs. It is clear that without the LiF layer, electron injection in the device is not very efficient and similarly hole injection is bad when the PEDOT:PSS layer is absent. The magneto-transport properties of these three kind of devices show that upon poor electron or hole injection the MR response still remain close to zero and an order less than that of the diodes in Fig. 1, as expected from the normalized recombination ratio. However, it can be seen that there are some variations in the OMAR magnitude, when the carrier balance in the solar cells is disrupted. The OMAR turns negative for the hole-rich device and positive for the electron-rich device (Fig. 4(b)). If we compare Fig. 4(a) and (b) it is evident that at higher voltages the current in these devices (without PEDOT:PSS and LiF) show clear sign of saturation and MR starts increasing from that particular voltage value where the saturation begins, as indicated by arrows. The solid lines in Fig 4(a) shows the space charge limited current (SCLC) values calculated using Child's law,

$$j = \frac{9}{8}\varepsilon\varepsilon_0(\mu_n + \mu_p)\frac{V^2}{d^3} \qquad (2)$$

where $\varepsilon$ is permittivity, $\mu_n$ and $\mu_p$ are mobilities of holes and electrons, respectively and d is the thickness of the device, as expected in single carrier devices or in bipolar devices with fast (Langevin) recombination. We used a value of $5 \times 10^{-3}$ cm$^2$/Vs for the mobility. The variations observed are due to different thickness and device areas. It becomes clear from this plot that the devices with unbalanced charge injection reach the unipolar SCLC region at the same time as MR turns negative(positive) for hole(electron) dominated devices. In the good solar cells the high current values are a direct consequence of plasma formation of charge carriers in the material due to reduced recombination [17] and the MR is always zero.

In conclusion, we have used BHSC to clarify the role of formation of coulombically bound e-h pairs on the MR effect. BHSCs are good model system since the excitons are effectively quenched. By systematically varying the probability for e-h pairs to meet in space, as observed by the experimentally measured recombination coefficient, we found that a reduced probability for electrons and holes to meet in space lead to a vanishing MR. The MR increases with increasing probability of e-h pair formation, as observed in BHSCs from MDMO-PPV:PCBM blends. We also found that the OMAR line shape in RRP3HT diodes follow a power law-dependence of $B^{0.5}$ at higher fields, deviating from both the Lorentzian and non-Lorentzian line-shape proposed in the bipolaron model [5].

The authors acknowledge Prof. M. Wohlgenannt and Prof. P. Bobbert for valuable discussions, G. Sliauzys for helping with the measurements and financial support from Academy of Finland for funding from project 116995. S.M. acknowledges the SITRA-CIMO fellowship. Planar International Ltd. is acknowledged for the patterned ITO substrates.

**Figure captions:**

**Figure 1. (a)** % MR as a function of magnetic field of a typical RRP3HT diode device measured with 10, 20, 50 and 100 µA currents at room temperature. **(b)** % MR as a function of magnetic field for the same diode device for the same currents showing √B dependence at higher magnetic fields with a deviation in lower magnetic field and lower driving currents.

**Figure 2.** *I-V* characteristics (black) and %MR (red) as a function of voltage in a typical RRP3HT diode (filled symbols) and a RRP3HT:PCBM solar cell (open symbol) showing a reduced probability of forming e–h pairs.

**Figure 3.** %MR as a function of magnetic field (B) in different devices with varying $\beta/\beta_L$ ratio. (●) is a RRP3HT diode, (o) is RRP3HT:PCBM BHSCs made from dicholorobenzene, (▲) is RRP3HT:PCBM BHSCs made from chloroform and (∆) is MDMO-PPV:PCBM BHSCs.

**Figure 4.** (a) *I-V* characteristics of RRP3HT:PCBM BHSCs with different electrode conditions. The solid lines are SCLC current calculated using Eq. 2. (b) for the three kinds of devices. The solid and the dashed arrows are only guides to the eye showing that MR starts increasing at the same voltage where SCLC conduction starts in the unbalanced devices. The inset shows the energy band diagram of a typical BHSC device.

**Figures:**

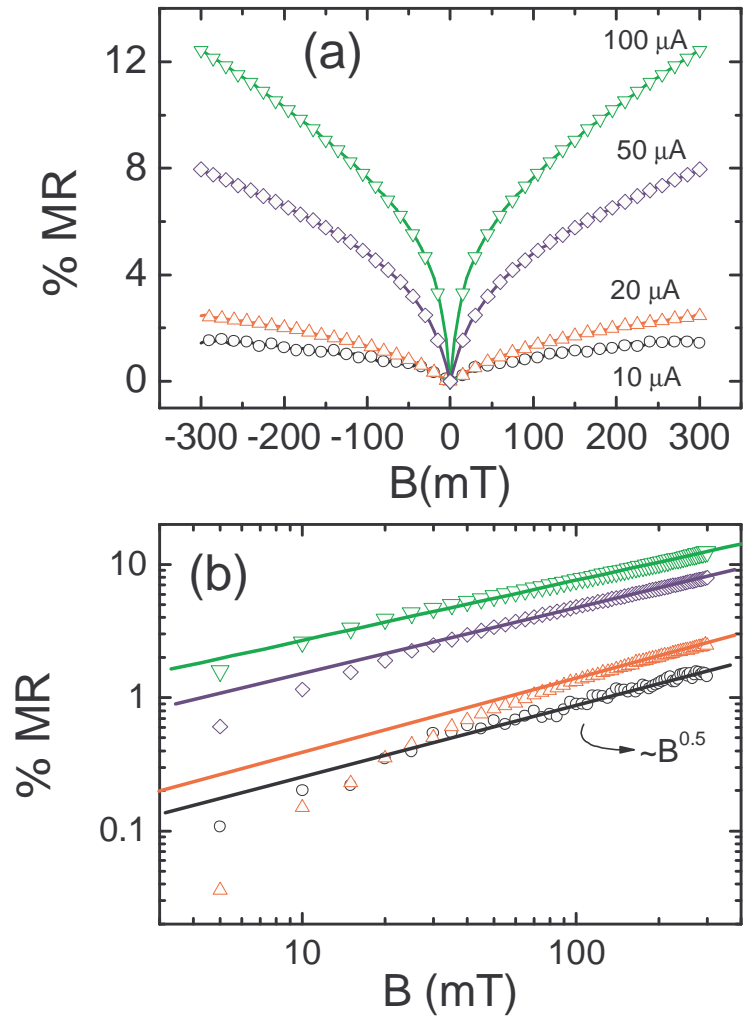

Figure 1. Sayani Majumdar et. al.

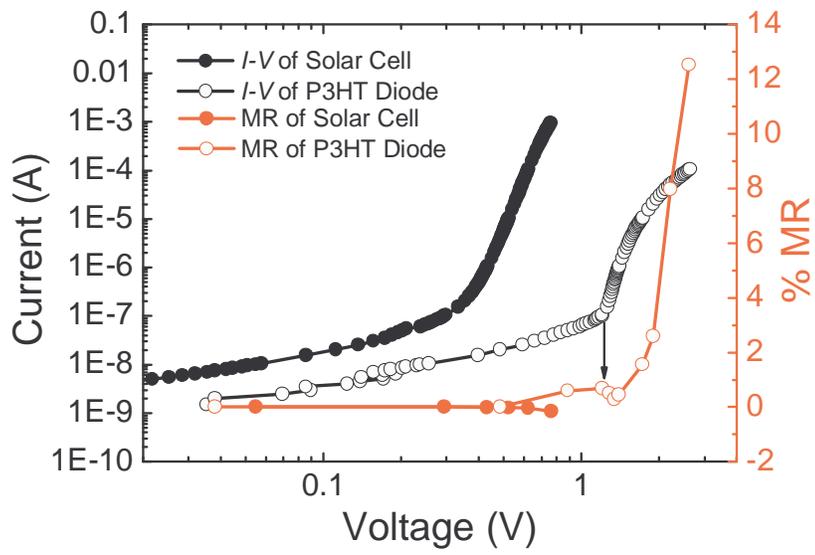

Figure 2. Sayani Majumdar et. al.

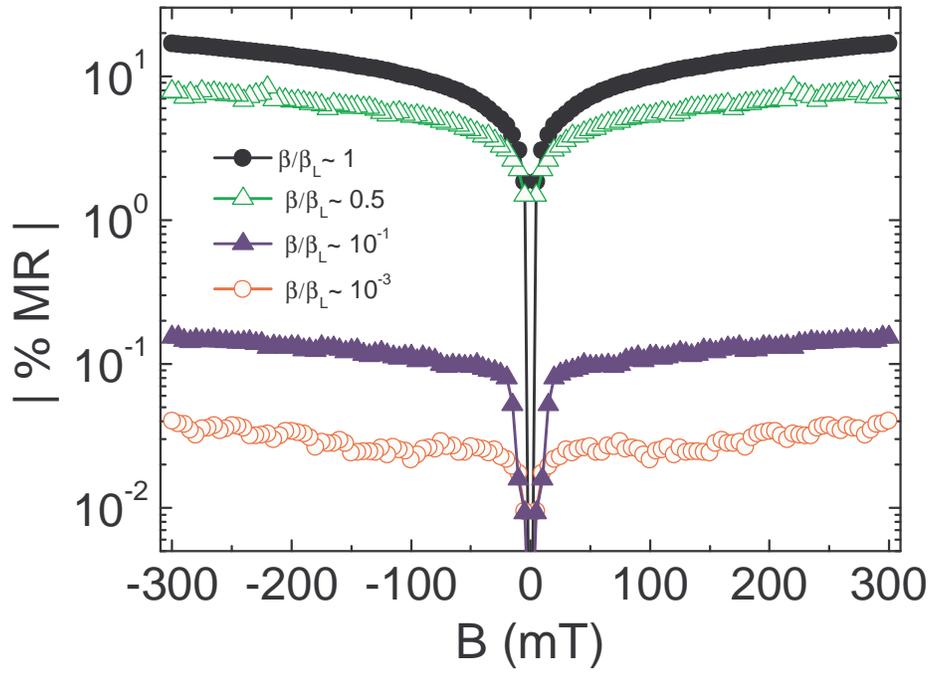

Figure 3. Sayani Majumdar et. al.

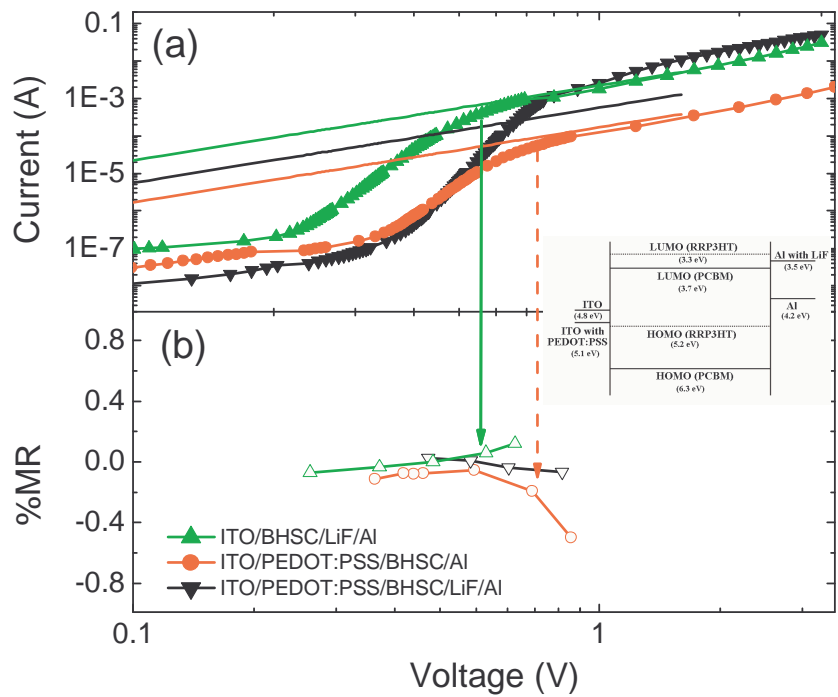

Figure 4. Sayani Majumdar et. al.